\newif\ifonecolumn
\onecolumntrue 
\onecolumnfalse
\ifonecolumn
  \documentclass[journal,draftcls,onecolumn, 12pt]{IEEEtran}
\else
  \documentclass[journal]{IEEEtran}
\fi

\usepackage[T1]{fontenc}

\usepackage{amsmath}
\usepackage{cite}
\usepackage{array}
\usepackage{booktabs}

\usepackage{stfloats}
\usepackage{url}
\usepackage{amsmath,amssymb}
\usepackage{epsfig,fancyhdr}
\usepackage{CJKutf8}
\usepackage{times}
\usepackage{graphicx}
\usepackage{multirow}
\usepackage{stfloats}
\usepackage{color}
\usepackage[dvipsnames]{xcolor}
\usepackage{soul}
\usepackage[normalem]{ulem}
\usepackage{balance}

\usepackage{bm}
\usepackage{diagbox}
\usepackage{threeparttable}
\usepackage{tablefootnote}

\newtheorem{thm}{Theorem}

\newtheorem{defin}{Definition}
\newtheorem{lemma}{Lemma}

\newtheorem{eg}{Example}
\newtheorem{rmk}{Remark}
\begin{document}
\title{Set Size Bound for Aperiodic Z-Complementary Sets}

\author{
Cheng-Yu~Pai,~\IEEEmembership{Member,~IEEE}, Yu-Che~Tung, Zhen-Ming~Huang,~\IEEEmembership{Graduate Student Member,~IEEE}, and Chao-Yu~Chen,~\IEEEmembership{Senior Member,~IEEE}

\thanks{Cheng-Yu Pai and Chao-Yu Chen are with the Department of Electrical Engineering and the Institute of Computer and Communication Engineering, National Cheng Kung University, Tainan 701, Taiwan (e-mail: cy$\_$pai@gs.ncku.edu.tw and super@mail.ncku.edu.tw). 

Yu-Che~Tung is with the Institute of Computer and Communication Engineering, National Cheng Kung University, Tainan 701, Taiwan (email:q36114132@gs.ncku.edu.tw).

Z.-M. Huang is with the Institute of Computer and Communication Engineering and the Department of Engineering Science,  National Cheng Kung University, Tainan 701, Taiwan (e-mail: n98101012@gs.ncku.edu.tw).}
}
\maketitle \begin{abstract}
The widely and commonly adopted upper bound on the set size of aperiodic Z-complementary sets (ZCSs) in the literature has been a conjecture. In this letter, we provide detailed derivations for this conjectured bound. A ZCS is optimal when its set size reaches the upper bound. Furthermore, we propose a new construction of ZCSs based on extended generalized Boolean functions (EGBFs). The proposed method introduces optimal ZCSs with new parameters.
\end{abstract}
\section{Introduction}
The Golay complementary set (GCS) was first introduced in \cite{Tseng}, where the constituent sequences exhibit the zero aperiodic autocorrelation sum property for all non-zero time shifts. From the application standpoint, GCSs have been widely applied in various fields, such as radar \cite{radar}, channel estimation \cite{channel}, synchronization \cite{CS_sync}, and reduction of peak-to-average power ratio (PAPR) in orthogonal frequency division multiplexing (OFDM) systems \cite{OFDM1,OFDM2,OFDM3,OFDM4}. 

The complete complementary code (CCC) \cite{CCC}, which is a collection of mutually orthogonal GCSs, plays a critical role in mitigating both multiple access interference (MAI) and multipath interference (MPI) in multicarrier code division multiple access (MC-CDMA) systems \cite{Chen_CDMA}. However, CCCs have a limitation in supporting a large number of users, as the set size (i.e., the number of users) is constrained by the number of sequences in a constituent GCS.
To address this issue, Fan {\it et al.} \cite{ZCP-1st} introduced the Z-complementary set (ZCS), which relaxes aperiodic auto- and cross-correlation constraints, allowing for more flexible set sizes compared to CCCs. In \cite{ZCP-1st}, a conjectured upper bound on the set size of ZCSs was provided as well.
 
The conjectured bound was modified by Feng {\it et al.} in \cite{ZCS_bound_Feng}, but it remains unproven to date. Additionally, \cite{ZCS_bound_Feng} presented another theoretical upper bound on the set size of ZCSs using the inner product theorem \cite{Welch}; however, this bound is not tight. Therefore, the widely referenced bound in the literature is still a conjectured bound, even to the extent that the optimal ZCS is defined according to the conjectured bound \cite{Wu_18,con_ZCS1,con_ZCS2,Sarkar_PBF,con_ZCS3,CCC1,Shen_EBF,Wu_21,Men_22,Liu_ZCS_23}. In this letter, we provide a detailed proof for this conjecture. Furthermore, we propose an algebraic construction of ZCSs based on extended generalized Boolean functions (EGBFs). Compared with current state-of-the-art constructions of ZCSs, the proposed method allows for new parameters, where the set size exactly meets the proposed upper bound with equality.


\section{Preliminaries and Definitions}\label{sec:background}
Throughout this letter, the following notations will be used:
\begin{itemize}
\item $\mathbb{Z}_q=\{0,1,\ldots,q-1\}$ represents the set of integers modulo a positive integer $q$;
\item $\xi=e^{2\pi \sqrt{-1}/q}$ is a primitive $q$-th root of unity;
\item we consider $q$ to be an even integer;
\item $(\cdot)^{*}$ denotes the complex conjugation;
\item $\lfloor \cdot\rfloor$ denotes the greatest integer less than or equal to the given value.
\end{itemize}
Let ${\bm c=(c_0,c_1,\ldots,c_{L-1})}$ and ${\bm d=(d_0,d_1,\ldots,d_{L-1})}$ be two unimodular sequences of length $L$. Then, the {\em aperiodic cross-correlation function} (ACCF) of $\bm c$ and $\bm d$ at shift $u$ is defined as
\begin{equation}
\begin{aligned}
{\rho}({\bm c},{\bm d};u)=\begin{cases}
\sum\limits_{i=0}^{L-1-u}{c_{i+u}d^*_i},\quad 0 \leq u\leq L-1;\\
\sum\limits_{i=0}^{L-1+u}{c_{i}d^*_{i-u}},\quad -L+1 \leq u<0.
\end{cases}
\end{aligned}
\label{eq:ACCF}
\end{equation}
When $\bm c=\bm d$, ${\rho}(\bm c,\bm c;u)$ can be simply expressed as ${\rho}(\bm c;u)$ and referred to as the {\em aperiodic autocorrelation function} (AACF) of $\bm c$.
\begin{defin}\label{def:ZCS}
 Given a set of $M$ sequence sets  ${{C}} = \{C^{0},C^{1},\ldots,C^{M-1}\}$ where each constituent set is composed of $N$ unimodular sequences with each sequence having a length of $L$, i.e., $C^{p} = \{\bm{c}_{0}^{p},\bm{c}_{1}^{p},\ldots,\bm{c}_{N-1}^{p}\}$ for $p=0,1,\ldots,M-1$. The set $C$ is called an $(M,N,L,Z)$-ZCS, if
    \begin{equation} \label{ZCS_def}
		\begin{aligned}
			{\rho}(C^{p},C^{t};u) &\triangleq \sum_{\lambda=0}^{N-1}{\rho}({\bm c}^{p}_{\lambda},{\bm
            c}^{t}_{\lambda};u)
            \\ &=
            \begin{cases}
                NL, & u=0,  \hspace{0.2cm} p=t; \\
                0,  & 0 < |u| < Z,\ p=t;\\
                0,  & |u| < Z,\ p \neq t.\\
            \end{cases} \\
		\end{aligned}
	\end{equation}
    \end{defin}
    where $M, N, L,$ and $Z$ represent the set size, flock size, sequence length, and the width of ZCZ width, respectively. If  $M = N$ and $Z = L$, the set is reduced to a {\em complete complementary code}, denoted by $(M,M,L)$-CCC. 
    
    It is well known that the set size of the CCC is at most equal to the flock size \cite{CCC}, whereas the size of the ZCS can be larger than the flock size.
    
In \cite{ZCP-1st}, the upper bound on the set size of $(M,N,L,Z)$-ZCSs was first conjectured as 
\begin{equation}\label{eq:Fan_conjecture}
M\leq N\left \lfloor\frac{L}{Z}\right\rfloor.
\end{equation}    
Later in \cite{ZCS_bound_Feng}, the conjectured bound was modified as 
\begin{equation}\label{eq:Feng_conj_bound}
M\leq \left \lfloor\frac{NL}{Z}\right\rfloor.
\end{equation} 
Note that the bound in (\ref{eq:Feng_conj_bound}) has been proven for periodic ZCSs in \cite{ZCS_bound_Feng}, but it is still a conjectured bound for aperiodic ZCSs.
Moreover, \cite{ZCS_bound_Feng} provided another theoretical upper bound on the set size of the $(M,N,L,Z)$-ZCS as given by
\begin{equation}\label{eq:Feng_bound}
\begin{aligned}
M \leq \frac{N (L+Z-1)}{Z}
\end{aligned}
\end{equation}    
using the inner product theorem from \cite{Welch}.
However, it is much looser compared to the upper bound in (\ref{eq:Feng_conj_bound}). Therefore, even though the conjectured bounds in (\ref{eq:Fan_conjecture}) and (\ref{eq:Feng_conj_bound}) have not been proven, they have been used in defining the optimal ZCS \cite{Wu_18,con_ZCS1,con_ZCS2,Sarkar_PBF,con_ZCS3,CCC1,Shen_EBF,Wu_21,Men_22,Liu_ZCS_23}.

\subsection{Generalized Boolean function}
Given a GBF $f:\mathbb{Z}^{m}_2\rightarrow \mathbb{Z}_q$ that consists of $m$ variables $(x_1,x_2,\ldots,x_m)\in \mathbb{Z}^m_2$ \cite{Paterson00}, the product of $r$ distinct variables is called a monomial of degree $r$. For instance, the term $x_1x_3$ represents a monomial of degree $2$.
The sequence $\bm f$ associated with a $q$-ary GBF $f$ is expressed as ${\bm f} = (f_0, f_1, \ldots, f_{2^m-1})$, where $f_i = f(i_1, i_2, \ldots, i_m)$ and $i = \sum_{l=1}^{m} i_l 2^{l-1}$. The corresponding complex-valued sequence, denoted as $\phi({\bm f})$, is represented by 
\begin{equation}
\phi({\bm f})=\left(\xi^{f_0},\xi^{f_1},\ldots,\xi^{f_{2^{m}-1}}\right).
\end{equation}
To construct sequences of lengths not restricted to powers of 2, define the truncated sequence ${\bm f}^{(L)}$ by omitting the last $2^{m}-L$ entries from ${\bm f}$, i.e., ${\bm f}^{(L)}=(f_0,f_1,\ldots,f_{L-1})$. For instance, given $m=3$, $q=4$, and $L=5$, the sequence associated with the GBF $f=3x_1$ is ${\bm f}=3{\bm x}_1=(03030303)$, and the truncated sequence is ${\bm f}^{(5)}=3{\bm x}^{(5)}_1=(03030)$.
\subsection{Extended Generalized Boolean function} \label{sec:EGBF}
An EGBF \cite{Pai_23} of $n$ variables is a function mapping $g:(y_1,y_2,\ldots,y_n)\in\mathbb{Z}^{n}_b\rightarrow g(y_1,y_2,\ldots,y_n)\in\mathbb{Z}_q$ where $y_l\in \mathbb{Z}_b$ for $l=1,2,\ldots,n$ and $2\leq b\leq q$. The sequence ${\bm g}$ associated with the EGBF $g$ is ${\bm g}=(g_0, g_1, \ldots, g_{b^n-1})$ with $g_h=g(h_1,h_2,\ldots,h_n)$ and $h = \sum_{l=1}^{n} h_l b^{l-1}$. For example, taking $n=2$, $b=3$, and $q=3$, the sequences corresponding to the EGBFs $y_1$ and $y_2$ are ${\bm y}_1=(012012012)$ and ${\bm y}_2=(000111222)$, respectively. When $b=2$, an EGBF reduces to a GBF and when $b=q$, it becomes an extended Boolean function (EBF).

\section{Derivation of the Upper Bound on the Set Size of ZCSs}\label{sec:proof}
First, we present two essential lemmas.
\begin{lemma}\cite{Welch}\label{lemma:welch}
A matrix $\bm X$ of size $\overline L\times \overline M$ can be represented by
\begin{equation}\label{eq:welch_matrix}
\bm X \triangleq \left[ \bm s^T_0, \bm s^T_1,\ldots, \bm s^T_{\overline M-1} \right],
\end{equation} 
where $\bm s_v=( s_{0, v}, s_{1, v},\ldots, s_{\overline L-1, v})$ and $\sum\limits_{l=0}^{\overline L-1} |s_{l,v}|^2=E$, for $v=0,1,\ldots,\overline M-1$. Let $\delta_{\text{max}}=\max\limits_{v \neq t} \left| \sum\limits_{l=0}^{\overline L-1} s_{l,v} s_{l,t}^*\right|$ be the maximum value of the inner product between any two distinct column vectors in ${\bm X}$. Then we have
\begin{equation} \label{eq:welch}
    \begin{aligned}
    \overline M(\overline M-1)\delta_{\mathop{\max}}^{2}+\overline M\cdot (E)^{2} 
    &\ge \sum\limits_{v=0}^{\overline M-1} \sum\limits_{t=0}^{\overline M-1}\left|  \sum\limits_{l=0}^{\overline L-1} s_{l,v} s_{l,t}^*\right|^{2}\\
    &=\sum\limits_{l=0}^{\overline L-1} \sum\limits_{l'=0}^{\overline L-1}\left|  \sum\limits_{v=0}^{\overline M-1} s_{l,v} s_{l',v}^*\right|^{2}.
    \end{aligned}
    \end{equation}
\end{lemma}
\begin{lemma}\label{lemma:property}
Given two orthogonal sequences $\bm c=(c_0,c_1,...,c_{\overline L-1})$ and $\bm d=(d_0,d_1,...,d_{\overline L-1}  )$ where $|c_{\alpha}|=|d_{\alpha}|=1$ for $0\leq \alpha\leq \overline L-1$, we have
    \begin{equation}
    \begin{aligned}
    \left |\sum_{\substack{l=0\\ l \neq \alpha}}^{\overline L-1}c_l d_l^*\right |^2 = \left |-c_\alpha d_\alpha^*\right |^2=1.
    \end{aligned}
    \end{equation}
\end{lemma}
\begin{IEEEproof}
Since $\bm c$ and $\bm d$ are mutually orthogonal, i.e., $\sum\limits_{l=0}^{\overline L-1}c_l d_l^* =0$,
we obtain the equality
$\sum\limits_{l=0, l \neq \alpha}^{\overline L-1}c_l d_l^* = -c_\alpha d_\alpha^*$.
Therefore, $\left |\sum\limits_{l=0, l \neq \alpha}^{\overline L-1}c_l d_l^*\right |^2 = \left |-c_\alpha d_\alpha^*\right |^2=1$, which complete the proof.
\end{IEEEproof}

Using the two preceding lemmas, we could prove the conjectured upper bound on the set size of ZCSs in (\ref{eq:Feng_conj_bound}).
\begin{thm}\label{thm:bound}
 Given an $(M, N, L, Z)$-ZCS, the upper bound on set size is given by
    \begin{equation} \label{eq:ZCS_bound}
    \begin{aligned}
    M \le \left \lfloor \frac{NL}{Z} \right \rfloor.
    \end{aligned}
    \end{equation}
\end{thm}
\begin{IEEEproof}
First, define $S^u({\bm c})$ as the $u$-th right cyclic shift of the sequence $\bm c=(c_0,c_1,\ldots,c_{\overline L-1})$, i.e., 
\begin{equation}
S^u (\bm c) \triangleq (c_{\overline L-u}, c_{\overline L-u+1}, \ldots, c_{\overline L-1}, c_{0},\ldots, c_{\overline L-u-1}).
\end{equation}
For an $(M, N, L, Z)$-ZCS $C=\{C^0,C^1,\ldots,C^{M-1}\}$ as defined in {\it Definition \ref{def:ZCS}}, let $\overline M=MZ$ and $\overline L=N(L + Z - 1)$. Then let the matrix 
 \begin{equation}\label{eq:matrix_ZCS}
\left[ \bm s^T_0, \bm s^T_1,\ldots, \bm s^T_{\overline M-1} \right]_{\overline L\times \overline M},
 \end{equation}
 where $\bm s_{(pZ+u)}=S^u\left((\bm{c}^p_0, \bm 0_{Z-1}, \bm{c}^p_1, \bm 0_{Z-1}, \ldots, \bm{c}^p_{N-1}, \bm 0_{Z-1})\right)$ and ${\bm 0}_{Z-1}$ denotes an all-zero vector of length $Z-1$ for $p=0,1,\ldots,M-1$ and $u=0,1,\ldots,Z-1$. Note that (\ref{eq:matrix_ZCS}) has the same form as (\ref{eq:welch_matrix}) and every column of (\ref{eq:matrix_ZCS}) has the same energy $NL$.
Since $u\leq L-1$, there exists some $\alpha\in \mathbb{Z}_{\overline L}$ such that $|s_{\alpha,v}|=1$ for all $v=0,1,\ldots,\overline M-1$. Then, the summations on the right-hand side of the second equality in (\ref{eq:welch}) can be further expressed as
\begin{equation}\label{eq:ZCS_equality}
\begin{aligned}
&\sum\limits_{l=0}^{\overline L-1} \sum\limits_{l'=0}^{\overline L-1}\left|  \sum\limits_{v=0}^{\overline M-1} s_{l,v} s_{l',v}^*\right|^{2}\\
=&\sum_{\substack{l=0\\ l\neq \alpha}}^{\overline L-1} \sum_{\substack{l'=0\\l'\neq \alpha}}^{\overline L-1}  \left | \sum_{v=0}^{\overline M-1}s_{l,v} s_{l',v}^* \right |^2 + \sum_{\substack{l=0\\ l \neq \alpha}}^{\overline L-1}  \left | \sum_{v=0}^{\overline M-1}s_{l,v} s_{\alpha,v}^* \right |^2 \\
&+\sum_{\substack{l'=0\\ l' \neq \alpha}}^{\overline L-1}  \left | \sum_{v=0}^{\overline M-1}s_{\alpha,v} s_{l',v}^* \right |^2+\left | \sum_{v=0}^{\overline M-1}s_{\alpha,v} s_{\alpha,v}^* \right |^2.
\end{aligned}
\end{equation}
Through some manipulations of (\ref{eq:ZCS_equality}), we obtain 
\begin{equation}
\begin{aligned}
&\sum\limits_{l=0}^{\overline L-1} \sum\limits_{l'=0}^{\overline L-1}\left|  \sum\limits_{v=0}^{\overline M-1} s_{l,v} s_{l',v}^*\right|^{2}-\sum_{\substack{l=0\\ l\neq \alpha}}^{\overline L-1} \sum_{\substack{l'=0\\l'\neq \alpha}}^{\overline L-1}  \left | \sum_{v=0}^{\overline M-1}s_{l,v} s_{l',v}^* \right |^2\\
&-\left | \sum_{v=0}^{\overline M-1}s_{\alpha,v} s_{\alpha,v}^* \right |^2= 2\sum_{\substack{l=0\\ l \neq \alpha}}^{\overline L-1}  \left | \sum_{v=0}^{\overline M-1}s_{l,v} s_{\alpha,v}^* \right |^2 \geq 0.
\end{aligned}
\end{equation}
Subsequently, interchanging the order of summation leads to
\begin{equation}\label{eq:ZCS_equality2}
\begin{aligned}
&\sum_{v=0}^{\overline M-1} \sum_{t=0}^{\overline M-1}  \left | \sum_{l=0}^{\overline L-1}s_{l,v} s_{l,t}^* \right |^2-\sum_{v=0}^{\overline M-1} \sum_{t=0}^{\overline M-1}  \left | \sum_{\substack{l=0\\ l \neq \alpha}}^{\overline L-1}s_{l,v} s_{l,t}^* \right |^2  \\&- \left | \sum_{v=0}^{\overline M-1}s_{\alpha,v} s_{\alpha,v}^* \right |^2 \ge 0. \\
\end{aligned}
\end{equation}
We have
\begin{equation}
\sum\limits_{l=0}^{\overline L-1} s_{l,v}s_{l,t}^*= 
\begin{cases}
                NL, & v=t; \\
                0,  & v\neq t,
 \end{cases} 
\end{equation}
due to the properties of the ZCS. Additionally, $\left | \sum\limits_{\substack{l=0\\ l \neq \alpha}}^{\overline L-1}s_{l,v} s_{l,t}^* \right |^2=1$ for  $0\leq v\neq t\leq \overline L-1$ based on {\it Lemma~\ref{lemma:property}}, (\ref{eq:ZCS_equality2}) can be rewritten as  
\begin{equation}
\begin{aligned}
\overline M\cdot (NL)^{2}  - \overline M\cdot (NL-1)^{2} - \overline M(\overline M-1) \cdot 1 - \overline M^2 \geq 0
\end{aligned}
\end{equation}
which is equivalent to
\begin{equation}
MNLZ-(MZ)^2\geq 0
\end{equation}
by substituting $\overline M$ as $MZ$. Since the set size $M$ is an integer, we possess
\begin{equation}
 M \leq \left \lfloor \frac{NL}{Z} \right \rfloor,
\end{equation}
which completes the proof.
\end{IEEEproof}
\begin{defin}\label{def:optimal}
An $(M,N,L,Z)$-ZCS is referred to as optimal if the equality of (\ref{eq:ZCS_bound}) holds.
\end{defin}
\begin{rmk}
{\it Theorem \ref{thm:bound}} confirms the correctness of the conjectured upper bound on the set size of ZCSs as presented in \cite{ZCS_bound_Feng}.
\end{rmk}
\section{Proposed Construction of ZCSs}\label{sec:ZCS}
In this section, we present a novel construction of ZCSs using the technique of EGBFs.

\begin{thm}\label{thm:ZCS}
For non-negative integers $m$ and $k$, where $k\leq m$, define the nonempty sets $W_1,W_2,\ldots W_k$ as a partition of the set $\{1,2,\ldots,m\}$. Then let $\pi_{\gamma}$ be a bijection from $\{1,2,\ldots,m_{\gamma}\}$ to $W_{\gamma}$ for $\gamma=1,2,\ldots,k$, where $m=\sum\limits_{\gamma=1}^{k}m_{\gamma}$ and $m_{\gamma}=|W_{\gamma}|$. Let a GBF $f:\mathbb{Z}^{m}_2\rightarrow \mathbb{Z}_q$ be
\begin{equation}
f=\frac{q}{2}\sum_{\gamma=1}^{k}\sum_{l=1}^{m_{\gamma}-1}x_{\pi_{\gamma}(l)}x_{\pi_{\gamma}(l+1)}+\sum_{l=1}^{m}\beta_{l}x_{l}+\beta_0
\end{equation}
where $\beta_l\in \mathbb{Z}_q$ for $l=0,1,\ldots,m$. Choose an integer $b$ such that $b|q$ and $b^n\leq 2^m$. Then, define EGBFs $g^p:\mathbb{Z}^{n}_b\rightarrow \mathbb{Z}_q$ for $p=0,1,\ldots,b^n-1$ by
\begin{equation}
g^p=\frac{q}{b}\sum_{l=1}^{n}p_{l}y_l
\end{equation}
where $(p_1,p_2,\ldots,p_n)$ is the base-$b$ representation of the integer $p$.
If $\{\pi_1(1),\pi_2(1),\ldots,\pi_k(1)\}=\{1,2,\ldots,k\}$, let
\begin{equation}
C^p=\left\{\phi\left({\bm f}^{(b^n)}+{\bm g}^p+\frac{q}{2}\sum_{\gamma=1}^{k}\lambda_{\gamma}{\bm x}^{(b^n)}_{\pi_{\gamma}(1)}\right):\lambda_{\gamma}\in \mathbb{Z}_{2}\right\}.
\end{equation}
Then, the set $\{C^0,C^1,\ldots,C^{b^n-1}\}$ forms a $(b^n,2^k,b^n,2^k)$-ZCS.
\end{thm}
\begin{IEEEproof}
The proof is provided in the Appendix.
\end{IEEEproof}
\begin{rmk}\label{rmk:ZCS}
The constructed $(b^n,2^k,b^n,2^k)$-ZCS from {\it Theorem~\ref{thm:ZCS}} is optimal since $b^n=({2^kb^n})/{2^k}$
according to {\it Definition~\ref{def:optimal}}.
\end{rmk}
\begin{eg}
For $m=3$, $n=1$, $k=2$, $b=6$, and $q=6$, let $W_1=\{1,3\}$ and $W_2=\{2\}$ with $\pi_{1}=(1,3)$ and $\pi_{2}=(2)$. The function $f$ is $f=3x_1x_3$ where $\beta_l=0$ for $l=0,1,2,3$ and functions $g^p=p_1y_1$ for $p=0,1,\ldots,5$. The set $\{C^0,C^1,\ldots,C^{5}\}$ forms an optimal $(6,4,6,4)$-ZCS, where
\begin{equation*}
C^p=\left\{\phi\left({\bm f}^{(6)}+{\bm g}^p+3\lambda_1{\bm x}^{(6)}_1+3\lambda_2{\bm x}^{(6)}_2\right):\lambda_{l}\in \mathbb{Z}_2\right\}
\end{equation*}
for $p=0,1,\ldots,5$. The constructed $(6,4,6,4)$-ZCS is presented in Table \ref{table_ZCS_eg}, where constituent values represent the exponent of $\xi=e^{2\pi \sqrt{-1}/6}$. To the best of our knowledge, this optimal $(6,4,6,4)$-ZCS has not been previously reported.
\end{eg}
    \begin{table}[ht]
        \centering
        \caption{An Optimal (6, 4, 6, 4)-ZCS} \label{table_ZCS_eg}
    \begin{tabular}{|c|c|c|c|c|c|}
    \hline
    $ C^0 = \left\{\begin{matrix} 000003 \\ 030300 \\ 003303 \\ 033000 \end{matrix}\right\}$ & $ C^1 = \left\{\begin{matrix} 012342\\042045\\015042\\045345
 \end{matrix}\right\}$ & $ C^2 = \left\{\begin{matrix} 024021\\054324\\021321\\051024 \end{matrix}\right\}$\\ \hline
    $ C^3 = \left\{\begin{matrix} 030300\\000003\\033000\\003303 \end{matrix}\right\}$ & $ C^4 = \left\{\begin{matrix} 042045\\012342\\045345\\015042 \end{matrix}\right\}$ & $ C^5 = \left\{\begin{matrix} 054324\\024021\\051024\\021321 \end{matrix}\right\}$\\ \hline
    \end{tabular}
    \end{table} 

\begin{figure}[htb]
        \centering
        \includegraphics[width=80mm]{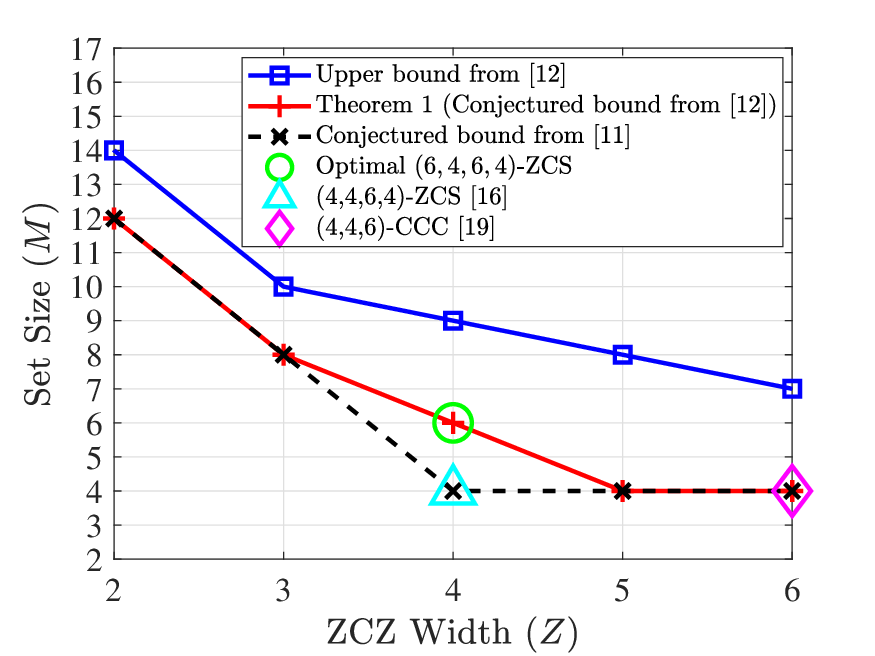}
        \caption{Comparison of known upper bounds on the set size of ZCSs with $N=4$ and $L=6$.} 
        \label{fig:bound}
    \end{figure}

Table \ref{table_ZCS} compares the existing ZCS constructions based on Boolean functions. It can be observed that {\it Theorem~\ref{thm:ZCS}} presents new parameters of optimal ZCSs that have never been previously reported, such as the $(6,4,6,4)$-ZCS and the $(9,4,9,4)$-ZCS. Furthermore, the set size of the proposed ZCSs attains the theoretical upper bound as stated in {\it Remark~{\ref{rmk:ZCS}}}.
\begin{table*}[htb]
\centering
\caption{Constructions of ZCSs Based on Boolean Functions}\label{table_ZCS}
\begin{tabular}{|c||c|c|c|c|c|c|}\hline
Method                                               & Parameters $(M,N,L,Z)$                & Based on                      & Note   &  Optimal ZCS                             \\ \hline
\cite{Wu_18}                                   &   $(2^{k+v},2^k,2^m,2^{m-v})$         & GBF  &  $m\geq 3$; $v\leq m$; $k\leq m-v$. & $\surd$    \\ \hline
\cite{con_ZCS2}                                   &   $(2^{k+1},2^{k+1},2^{m-1}+2,3\cdot 2^{m-3}+1)$        &  GBF  & $m\geq 3$.  &     \\ \hline
\cite{Sarkar_PBF}                                  &   $(v2^{k+1},2^{k+1},v2^m,2^m)$         & Pseudo Boolean function  & $m\geq 2$; $v$ is prime.  & $\surd$      \\ \hline
\cite{con_ZCS3}                                  &   $(2^{k+1},2^{k+1},3\cdot 2^m,2^{m+1})$   &  GBF  & $k\geq 1$; $m\geq 1$. &     \\ \hline
\cite{Shen_EBF}                      &   $(q^{k+1},q,q^m,q^{m-k})$         &  EBF  & $k\leq m$.   &  $\surd$  \\ \hline
{\it Theorem \ref{thm:ZCS}}          & $(b^n,2^k,b^n,2^k)$ & EGBF  & $b^n\leq 2^m$; $k\leq m$. & $\surd$
\\ \hline
\end{tabular}
\end{table*}
Fig. \ref{fig:bound} compares the upper bound from {\it Theorem~\ref{thm:bound}} (i.e., the conjectured bound from \cite{ZCS_bound_Feng}) with the conjectured bound from \cite{ZCP-1st}, and the theoretical bound from \cite{ZCS_bound_Feng} based on inner product theorem, corresponding to (\ref{eq:Fan_conjecture}) and (\ref{eq:Feng_bound}), respectively. For comparison, we set $N=4$ and $L=6$, and the parameters of existing ZCSs are also marked in Fig. \ref{fig:bound}. It is clear that the set size bound in {\it Theorem~\ref{thm:bound}} is tighter than the theoretical bound in \cite{ZCS_bound_Feng}. Additionally, the constructed $(6,4,6,4)$-ZCS from Table {\ref{table_ZCS_eg}} meets the upper bound specified in {\it Theorem \ref{thm:bound}}, i.e., $M=\left \lfloor(NL)/Z\right \rfloor$, confirming that this bound on the set size of ZCSs is both tight and achievable. Note that the conjectured upper bound from \cite{ZCP-1st} is not accurate since there exist ZCSs beyond this upper bound, e.g., the $(6,4,6,4)$-ZCS in Table \ref{table_ZCS_eg}.
\section{Conclusion}\label{sec:conclusion}
In the literature, there are two conjectured upper bounds on the set size of ZCSs, i.e., (\ref{eq:Fan_conjecture}) and (\ref{eq:Feng_conj_bound}). We have found that the upper bound in (\ref{eq:Fan_conjecture}) may not hold in some cases. Besides, we have provided a proof for the other conjectured bound in {\it Theorem~\ref{thm:bound}}. Therefore, the parameters of an optimal $(M,N,L,Z)$-ZCS should satisfy $M=\left \lfloor(NL)/Z\right \rfloor$. Moreover, by leveraging the technique of EGBFs, we introduce a construction of optimal ZCSs with new parameters in {\it Theorem \ref{thm:ZCS}}. 
\begin{appendix}[Proof of Theorem \ref{thm:ZCS}]\label{apxA}
Consider a ZCS $\{C^0,C^1,\ldots,C^{b^n-1}\}$ constructed from {\it Theorem~\ref{thm:bound}}, where $C^p=\{\phi({\bm c}^p_0),\phi({\bm c}^p_1),\ldots,\phi({\bm c}^p_{2^k-1})\}$ and ${\bm c}^p_\lambda$ is the $q$-ary sequence given by
\begin{equation}\label{eq:ZCS_form}
{\bm c}^p_\lambda={\bm f}^{(L)}+\frac{q}{b}\sum_{l=1}^{n}p_{l}{\bm y}_l+\frac{q}{2}\sum_{\gamma=1}^{k}\lambda_{\gamma}{\bm x}^{(L)}_{\pi_{\gamma}(1)},
\end{equation}
for $\lambda=0,1,\ldots,2^k-1$ and $L=b^n$. Hence, (\ref{ZCS_def}) is rewritten as
 \begin{equation}
		\begin{aligned}
			{\rho}(C^p,C^t;u)&=\sum_{\lambda=0}^{N-1}\sum_{i=0}^{b^n-1-u}\xi^{c^{p}_{\lambda,i+u}-c^{t}_{\lambda,i}}
            \\ &=
            \begin{cases}
                0,  & 0 < u < Z,\ p=t;\\
                0,  &  0\leq u < Z,\ p \neq t.\\
            \end{cases} \\
		\end{aligned}
	\end{equation}
Let $j=i+u$, and let $(j_1,j_2,\ldots,j_m)$ and $(i_1,i_2,\ldots,i_m)$ represent the binary forms of integers $j$ and $i$, respectively. Two cases are considered below.

{\it Case 1}: $0<u<Z$. We possess $i_{\pi_{\gamma}(1)}\neq j_{\pi_{\gamma}(1)}$ for some $\gamma\in \{1,2,\ldots,k\}$. Suppose not, i.e., $i_{\pi_{\gamma}(1)}= j_{\pi_{\gamma}(1)}$ for all $\gamma=1,2,\ldots,k$. Since $\{\pi_1(1),\pi_2(1),\ldots,\pi_k(1)\}=\{1,2,\ldots,k\}$, we have
\begin{equation}
\begin{aligned}
u=j-i&=\sum_{l=k+1}^{m}(j_l-i_l)2^{l-1}\geq 2^k,
\end{aligned}
\end{equation}
thereby contradicting the assumption of $u<2^k$. Hence, we have $i_{\pi_{\gamma}(1)}\neq j_{\pi_{\gamma}(1)}$ for some $\gamma\in \{1,2,\ldots,k\}$. For two sequences ${\bm c}^{p}_{\lambda}$ and ${\bm c}^{t}_{\lambda}$, there exist sequences ${\bm c}^{p}_{\lambda'}={\bm c}^{p}_{\lambda}+(q/2){\bm x}^{(L)}_{\pi_{\gamma}(1)}$ and ${\bm c}^{t}_{\lambda'}={\bm c}^{t}_{\lambda}+(q/2){\bm x}^{(L)}_{\pi_{\gamma}(1)}$, respectively, satisfying 
\begin{equation}
c^p_{\lambda,j}-c^{t}_{\lambda,i}-c^p_{\lambda',j}+c^t_{\lambda',i}= \frac{q}{2}(i_{\pi_{\gamma}(1)}-j_{\pi_{\gamma}(1)})\equiv \frac{q}{2} \pmod q,
\end{equation}
and thus $\xi^{c^p_{\lambda,j}-c^{t}_{\lambda,i}}+\xi^{c^p_{\lambda',j}+c^t_{\lambda',i}}=0$.

{\it Case 2}: $u=0$ and $p\neq t$. From (\ref{eq:ZCS_form}), we have
\begin{equation}\label{eq:ZCS_noshift}
{\bm c}^p_{\lambda}-{\bm c}^t_{\lambda}\equiv \frac{q}{b}\sum_{l=1}^{n}(p_{l}-t_l){\bm y}_l \pmod q,
\end{equation}
for $\lambda=0,1,\ldots,2^k-1$, where $(p_1,p_2,\ldots,p_n)$ and $(t_1,t_2,\ldots,t_n)$ are base-$b$ representations of $p$ and $t$, respectively. It can be seen that (\ref{eq:ZCS_noshift}) is the linear combination of sequences ${\bm y}_l$ associated with EGBFs $y_l$, implying that $\sum\limits_{i=0}^{b^n-1}\xi^{c^{p}_{\lambda,i}-c^{t}_{\lambda,i}}=0$ for all $\lambda=0,1,\ldots,2^k-1$. 

By combining Case 1 and Case 2, the proof is complete.
\end{appendix}

\balance
\bibliographystyle{IEEEtran}
\bibliography{IEEEabrv,ref}
\IEEEtriggeratref{3}
\end{document}